\documentstyle[11pt,jkas3]{article}
\begin{document}


\title{ THE ORIGIN OF LARGE SCALE GALACTIC MAGNETIC FIELDS}
\author{ K. SUBRAMANIAN  }

\address{NCRA - TIFR,
Poona University Campus, Ganeshkind, Pune 411007, India}

\abstract{Magnetic fields correlated on several kiloparsec scales are seen 
in spiral galaxies. Their origin could be due to the winding 
up of a primordial cosmological field or due to amplification 
of a small seed field by  a turbulent galactic dynamo. 
Both options have difficulties:
There is no known battery mechanism for producing the required primordial 
field. Equally the turbulent dynamo may self destruct 
before being able to produce the large scale field, due to excess 
generation of small scale power. 
The current status of these difficulties 
is discussed. The resolution could depend on the nature of the saturated
field produced by the small scale dynamo. We argue that the small scale
fields do not fill most of the volume of the fluid and 
instead concentrate into intermittent ropes, with their peak value
of order equipartition fields, and radii much smaller than their lengths.
In this case these fields neither drain significant 
energy from the turbulence nor convert eddy motion of the turbulence 
on the outer scale to wave like motion. This preserves the 
diffusive effects needed for the large scale dynamo operation. 
}


\maketitle  


\section{INTRODUCTION }

Magnetic fields in galaxies have strengths of order few $10^{-6} G$,
and are coherent on scales of several kpc (cf. Beck et al 1996).  
How such ordered, large scale fields arise is
a problem of considerable interest.
They can arise in principle,
due to dynamo amplification of a weak but nonzero seed field 
$\sim 10^{-19} - 10^{-23} G$,
if the galactic dynamo can operate efficiently to exponetiate the
field by a factor $\sim 30 - 40$ (cf.Zeldovich {\it et al.} 1983). 
But the origin of even such a small seed field
needs some physical explanation.  We review here some of the issues 
relevant to galactic magnetic field 
generation, in particular, 
the galactic dynamo theory,
its problems and possible solutions.           

The evolution of the magnetic field is generally 
described by the induction equation
$
(\partial {\bf B}/ \partial t) =
{\bf \nabla } \times ( {\bf v} \times {\bf B} - 
 \eta {\bf \nabla } \times {\bf B}), 
$ 
provided one assumes the usual form of Ohms law  and 
neglects the displacement current term in Maxwells equation.
Here ${\bf B}$ is the magnetic field,  
${\bf v}$ the velocity of the fluid and
$\eta$ the resistivity. If $\eta \to 0$ the magnetic flux through
any area in the fluid can be shown to be conserved during the
motion of the fluid. The presence of a finite resistivity allows
for a violation of such "flux freezing" and the magnetic reynolds
number $R_m = vL/\eta$ measures the relative 
importance of flux freezing versus resistive diffusion. 
(Here $v$ and $L$ are typical velocity and length scales of
the fluid motions.) In most astrophysical contexts 
flux freezing greatly dominates over diffusion with $R_m >> 1$.

Since ${\bf B} =0$ is a perfectly valid solution of the induction 
equation, there would be no magnetic fields generated if
one were to start with a zero magnetic field initially.
It is generally believed that the universe did not start with 
an initial magnetic field. So one needs
some way of violating the induction equation and produce a cosmic
battery effect, to drive curents from a state with initially
no current. There are a number of such battery mechanisms
which have been suggested (see Rees 1994; Subramanian 1995 for reviews).
All of them lead to only small fields much smaller
than the galactic fields. Therefore, it would be good to 
find velocity fields, which can act to exponentiate the small seed fields
efficiently. Some form of dynamo action is needed 
to explain the observed magnetic fields. We will
discuss below the possiblities of galactic dynamos after
touching briefly on one battery mechanism, which appears capable
of seeding the whole IGM with ordered fields, albiet with a small
value (Subramanian et al 1994). 

\section{A BATTERY MECHANISM}

The basic problem that any battery has to address is how to
produce finite currents from zero currents?
Most mechanisms use the fact that the positively and
negatively charged particles in a charge neutral universe do not 
have identical properties. For example if one considered a gas
of ionised hydrogen, then the electrons have a much smaller
mass compared to protons. This means that for a given pressure gradient
of the gas the electrons tend to be accelerated much more than the ions.
This leads in general to an electric field, which couples back
positive and negative charges, of the form
${\bf E}_T = {\bf \nabla}p_e / e n_e$, where $p_e$ and $n_e$ are
the electron pressure and number density, respectively. If such a
thermally generated electric field has a curl, then by 
Faradays law of induction a magnetic field can grow.
Taking $p_e = n_e kT$ with $T$ the electron temperature we have
${\bf \nabla} \times {\bf E}_T = - (c k/  e)
( {\bf \nabla } n_e / n_e)  \times {\bf \nabla } T$.
So ${\bf E}_T$ has a curl only if the density and
temperature gradients, are not parallel to each other. 
Biermann (1950) and Mestel and Roxburgh (1962)
applied this idea to stars. Subramanian et al (1994) have 
applied it to cosmic ionisation fronts, which are 
produced when the first UV sources turn on to
ionise the intergalactic medium.

The temperature gradient in a cosmic ionisation front is normal to the front.
However, a component to the
density gradient can arise in a different direction, if the ionisation
front is sweeping across arbitrarily laid down density
fluctuations, associated with protogalaxies/clusters since these
in general have no correlation to the source of the
ionising photons. The resulting thermally generated magnetic fields
on galactic scales turn out to have a strength $B \sim 3 \times 10^{-20} G$.
This field by itself is far short of the observed microgauss strength
fields in galaxies, but it can provide a seed field for a dynamo. 
Recently Kulsrud et al (1996) have suggested that the 
Biermann battery can also
operate in collapsing protogalaxies/clusters.

\section{ THE GALACTIC DYNAMO }

Spiral galaxies are differentially rotating systems.
Also the magnetic flux is to a 
large extent frozen into the fluid. So any radial
component of the magnetic field will be efficiently wound 
up and amplified to produce a toroidal component of the field.
But this results in only a linear amplification of the field 
and to obtain the observed 
galactic fields starting from small seed fields  
one should find a way to generate the radial 
components of the field in the galaxy from the toroidal one.
If this can be done, the field can grow exponentially and 
one has a dynamo.

A mechanism to produce the radial components from the toroidal 
field was originally invented by Parker (1955) (cf. 
Zeldovich et al 1983). The essential 
feature is to invoke the effects of cyclonic turbulence in the 
galactic gas. The galactic interstellar medium is assumed 
to be turbulent, due to for example the effect of supernovae randomly 
going off in different regions. In a rotating, stratified 
(in density and pressure) medium
like a disk galaxy, such turbulence becomes cyclonic and
aquires a net helicity. 
 Helical motions of the galactic 
gas perpendicular to the disk can draw out the toroidal field 
into a loop which looks like a twisted $\Omega$. Such a
loop is connected to a current and because of the twist this current has 
a component parallel to the original field. Since the gas has a
net helicity, in the presence of 
such motions a toroidal current can be produced from the toroidal field, 
Hence, poloidal fields can be generated from toroidal ones. 
In quantitative terms 
isotropic and homogeneous turbulence with helicity, in the 
presence of a large scale magnetic field {\bf B}, leads to an 
extra electromotive force of the form 
$ {\bf E} = \alpha {\bf B} - \eta_t {\bf \nabla } \times {\bf B}$ 
where $\alpha$ depends on the helical part of the turbulence 
and $\eta_t$ called the turbulent diffusion depends on the 
non helical part of the turbulent velocity correlation function.

A physics comment is in order at this stage. When one considers 
the effect of turbulent fluid motions on say smoke, one only gets
a mean diffusion of the smoke particles, associated with the 
random walking nature of turbulent fluid motions. But for magnetic 
fields the induction equation has terms which not only imply 
a body transport due to the random motions of the fluid, but also
a term which describes the generation of magnetic fields due to
velocity shear. It is this qualitative difference between magnetic
fields and smoke that leads to an alpha effect, over and
above turbulent diffusion (and also leads to the small scale dynamo
action discussed below). Note that both these effects also
crucially depend
on the diffusive (random walk) property of fluid motion. 
So if due to some reason (see below) the fluid motion becomes
wavelike, then the alpha effect and turbulent diffusion will
be suppressed. 

The induction equation, with the extra turbulent component of 
the electric field,  with a prescribed large scale velocity field, 
can have exponentially growing solutions for the large scale field.
These have been 
studied extensively in the literature (cf. Beck et al. 1996 for
a review). One can even modify it to discuss the 
possible reasons why large scale magnetic fields in
spirals are sometimes bi-symmetric and why these bi-symmetric
magnetic spirals are correlated with the optical spirals
(cf. Mestel and Subramanian 1991; Chiba and Tosa 1990).
We have assumed here that the turbulent 
velocities do not get affected by the Lorentz forces due to the 
magnetic field, at least not until the mean large scale 
field builds up sufficiently. However this does not turn out to
be valid due to the more rapid build up of magnetic noise 
compared to the mean field, a problem to which we now turn.

\section{PROBLEM OF MAGNETIC NOISE}

Suppose one splits up the magnetic field ${\bf B} = {\bf B}_0 
+ \delta{\bf B}$, into a mean field $ {\bf B}_0$ and a 
fluctuating component  $\delta{\bf B}$. Here the mean is defined
either as a spatial average over scales larger than the turbulent
eddy scales or more correctly as an ensemble average.
The dynamics of the fluctuating field has been 
worked out in detail by Kulsrud and Anderson (1992) (KA) in 
fourier space and by Subramanian (1996) using a complimentary
co-ordinate space approach. We summarise some of the results
drawing mainly on the later work. 

This analysis shows firstly that the fluctuating field, 
tangled on a scale $l$, can grow on the 
turn over time scale of a turbulent eddy of scale $l$, 
with a growth rate $\Gamma_l \sim v_l/l$, 
provided the magnetic reynolds number on that scale
$R_m(l) = v_l l/\eta $ is greater than a critical reynolds
number $R_c \sim 100$. Here $v_l$ is the velocity associated
with eddies of scale $l$. For Kolmogorov turbulence, since
$v_l \propto l^{1/3}$, $\Gamma_l \propto l^{-2/3}$.
If the magnetic reynolds number associated with eddies at the cut-off
scale (inner scale), say $l_c$,
of the turbulence is larger than $R_c$, then these 
eddies will themselves be able to exponentiate the magnetic
field first on these scales. Since the time scale for 
mean field growth is $\sim 10^9 yrs$, 
of order a few rotation time scales of the 
disk, is much larger than the turn around time scales of 
the turbulent eddies in the galaxy, the magnetic field 
will be rapidly dominated by the fluctuating component.
KA argued that as the small scale field builds up it will
drain energy from the turbulence, mainly due to the friction of the 
ionised gas and the neutrals resulting from ambipolar drift.
Also once the energy density 
in the small scale component achieves equipartition with 
the turbulent energy density, the turbulence will become
weak, a more wavelike "alfven" turbulence, than an eddy like
fluid turbulence resulting in a suppression of 
the alpha effect. All this
happens much before the mean field has grown 
appreciably. So KA speculated that the galactic field is 
primordial in origin.

\section{ A POSSIBLE SOLUTION }

Before accepting the above conclusions, it is worth 
re-examining carefully the dynamics of the small scale fields and 
its back reaction on the turbulence. We had pointed out earlier 
(Subramanian 1995) that if the small scale field is intermittent in 
space, then it may 
saturate without drastically draining the power from the turbulence.
This idea has been now investigated more 
thoroughly (Subramanian 1996), taking into account in a 
quantitative fashion the effects of ambipolar drift in the galaxy as well. 
We summarise below some of the relevant results.

For this it is useful look at the behaviour of the magnetic 
correlation function, say $w(r,t) = <\delta{\bf B}({\bf x},t).
\delta{\bf B}({\bf y},t)> $, where $r= \vert {\bf x} - {\bf y}\vert$ and
the angular brackets $<>$ indicates an ensemble average.
Suppose the turbulence was initially isotropic and 
homogeneous. Then in the kinematic regime, 
if $R_m(l_c) > R_c$, the fastest growing 
$w(r,t)$ has a form $f(r)e^{\Gamma t}$ with $f(r)$ strongly 
peaked within $r= r_1 = l_c / R_m^{1/2}(l_c)$, 
and a negative tail extending to $r_2 \sim l_c$. Zeldovich et al.
interpret such a correlation function as implying that field
is concentrated into ropes of thickness $ r_1$ and 
radius of order $r_2$. There can be 
slower growing higher order modes $w$ with more complicated structure
for the field but all with a ropy structure with rope
thickness of order $r_1$. The question arises as to 
how such ropes evolve in the non-linear regime when the lorentz force
due to the generated field reacts back on fluid motions?

We show that, in a partially ionised plasma, 
due to ambipolar drift, the effective diffusivity
changes to $\eta_{eff} = \eta +
<\delta{\bf B}^2> / ( 6\pi \rho_i \nu_{in})$, where $\rho_i$ is 
the ion density and $\nu_{in}$ is the neutral-ion collision 
frequency. So, as the energy density in the fluctuating field 
increases, the effective magnetic reynolds number, for
fluid motion on any scale of the turbulence
say $ R_{ambi}(l) = v_l l/ \eta_{eff}$, decreases. 
Firstly, this makes it easier for the field energy to reach the
diffusive scales $r_d \sim l/R_{ambi}^{1/2}$, from a general 
initial configuration. Subsequent field amplification, leads to a 
decreasing $R_{ambi}$ and an increase in $r_d$ and hence
the thickeness of the flux ropes. However for the galactic
turbulence with say an outer scale $L\sim 100pc$ and $v_L \sim 10$
km s$^{-1}$, $R_{ambi}(L) \sim n_if^{-1} 10^6 $, where $n_i$ is the
ion density in units of cm$^{-3}$, and $f$ is the ratio of 
the magnetic to the turbulent energy density $E_T$.
The flux ropes then remain relatively thin
with a thickness at most a value 
of order $L/R_{ambi}^{1/2} << L$, even taking account of 
the ambipolar drift.
The volume filling fraction of the field 
depends not only on
the thickness of the rope but also on its total length. 
In the kinematic regime, a rough estimate of the length
of the rope is $\sim N L$, where $N$ is related to the number
of higher order modes excited. For $R_{ambi} \sim 10^6$,
only a few higher order modes can be exited. In this case
the volume filling factor of the field is $\sim N/R_m$.
How this changes as the field reaches a stationary state is 
more difficult to quantify. (cf. Vishniac 1995, in a different context).

Since $R_{ambi} > R_c$, ambipolar drift alone cannot
lead to a saturation for the small scale dynamo.
The field keeps building up until the effects of the magnetic 
pressure in the ropes acting on the fluid as a whole
becomes important.
Due to the increasing importance of this pressure,
stretching of field lines can lead to a partial decrease in
fluid density in the ropes rather than a decrease
in the rope cross section and the associated increase
in the rope magnetic field. An upper limit to the
magnetic pressure in the ropes is given by the
external pressure $P_{ext}$. This implies that the field in the 
rope, say $B_r$ is limited to $B_r < (8\pi P_{ext})^{1/2}$.  
In the interstellar medium, the ratio of the gas pressure to the 
turbulent energy density $P_g/E_T \sim 1.7 (T/10^4 K)(v/10 km s^{-1})^{-2}$.
If $P_{ext}$ in the galaxy is dominated by the gas 
pressure, the peak field in the ropes will not be much larger than
the equipartition field. 
[Vishniac (1995) has also discussed the ropy nature of the magnetic
field in accretion disks. There are however crucial differences
due to the dominance of ambipolar drift in
the galaxy. Also in the case considered here, when the field 
grows from a small seed field, the kinematic evolution
naturally leads to an initial ropy structure for the field
which can be preserved when the field becomes strong]. 

The above considerations lead us to conjecture that when the small 
scale dynamo saturates, 1. The magnetic noise
generated by the small scale dynamo is ropy and does not
fill the volume of the fluid. 2. The peak fields in the 
ropes are not much greater than equipartition fields.
Given these conjectures one can show
that the power dissipated in ambipolar drift is much smaller 
than the turbulent power. Also, from the conjectures 1 and 2 
one can see that the average energy density of the 
generated small scale field is much smaller than the average
energy density in the turbulence. So any wave-like
motion induced by the presence of the field will have 
a period larger than the eddy turn around time. This implies
that such tangled small scale fields do not change the
diffusive nature of the turbulence. Due to the above reasons
the large scale dynamo can still operate to grow the mean 
field. More details of the above work can be found in 
Subramanian (1996). In conclusion, it seems that 
an understanding of how galactic magnetic fields originate, 
is far from complete and is still a challenging problem.

\acknowledgements
I thank the SOC/LOC of the Asia-Pacific IAU meeting for support.
Periodic discussions with T. Padmanabhan and a brief one
with Ethan Vishniac were very helpful.


\begin{references}
\reference {r2}

 

%
%
\reference{Beck96}
Beck, R. et al. 1996, \araa, (in Press).
\reference{Bier50}
Biermann, L., 1950. Zs. Naturforsch. A., 5, 65.
\reference{CT90}
Chiba M. $\&$ Tosa, M. 1990, \mnras, 244, 741.
\reference{ka92}
Kulsrud, R.M. $\&$ Anderson, S.W., 1992. ApJ., 396, 606.
\reference{Kulsrud96}
Kulsrud et al. 1996, (Submitted to ApJ, POPe-671).
\reference{Mestel62}
Mestel, L. $\&$ Roxburgh, I.W., 1962. ApJ., 136, 615.
\reference{Messub90}
Mestel, L. $\&$ Subramanian, K. 1991, \mnras, 248, 677.
\reference{Parker55}
Parker, E.N., 1955. ApJ., 122, 293.
\reference{Rees94}
Rees, M.J., 1994. {\it Cosmical Magnetism}, ed. Lynden-Bell, D., 
Kluwer, London, p155.
\reference{Subramanian95}
Subramanian, K. 1995, Bull.Astr.Soc. India., 23, 481.
\reference{Sub96}
Subramanian, K. 1996 (Preprint, to be submitted to MNRAS).
\reference{Subramanian94}
Subramanian, K., Narasimha, D. $\&$ Chitre, S.M., 1994. MNRAS. 271, L15.
\reference{vish95}
Vishniac, E. T. 1995, \apj, 446, 724.
\reference{Zel83}
Zeldovich, Ya.B., Ruzmaikin,A.A. $\&$ Sokoloff,D.D., 1983. {\it
Magnetic fields in Astrophysics}, Gordon and Breach, New York.
 
\end{references}
\end{document}